\documentstyle[12pt]{article}
\newcommand{\be}{\begin{equation}}
\newcommand{\ee}{\end{equation}}
\newcommand{\bea}{\begin{eqnarray}}
\newcommand{\eea}{\end{eqnarray}}
\newcommand{\half}{{\scriptstyle{{1\over 2}}}}
\newcommand{\quart}{{\scriptstyle{{1\over 4}}}}
\newcommand{\pf}{{\rm pf}}

\begin{document}
\hfill INLO-PUB-13/98
\vskip1cm
\begin{center}
{\LARGE{\bf{\underline{Nahm gauge fields for the torus}}}}\\
\vspace{7mm}
{\large Pierre van Baal} \\
\vspace{5mm}
Instituut-Lorentz for Theoretical Physics, University of Leiden,\\
PO Box 9506, NL-2300 RA Leiden, The Netherlands.
\end{center}
\vspace*{5mm}{\narrower\narrower{\noindent
\underline{Abstract:} We present the exact expression for the Nahm gauge field 
associated to a $SU(N)$ charge one self-dual gauge field on $T^3\times R$. The 
result implies that the size of the instanton is determined by the ``distance''
between its two flat connections at $t\rightarrow\pm\infty$.}\par}

\section*{}
1. A long standing problem is to find analytic self-dual solutions on a higher 
dimensional torus for non-Abelian gauge theories. For $T^4$ solutions with 
topological charges higher than one can be proven to exist~\cite{Tau}. However,
it can be shown there is no regular solution with topological charge 
one~\cite{BrvB}, though existence is assured~\cite{Bra} in case one allows 
for twisted boundary conditions~\cite{THotw}. The main tool for studying 
self-dual solutions has become the Nahm duality transformation~\cite{Nahm} 
that maps a charge $k$, $U(N)$ solution to a charge $N$, $U(k)$ solution on 
the dual torus. We consider gauge fields in four euclidean dimensions on spaces
$T^n\times R^{4-n}=R^4/\Lambda$, where $\Lambda$ is a $n$-dimensional lattice 
embedded in $R^4$, whose dual is denoted by the $n$-dimensional lattice 
$\hat\Lambda$, which we consider to be embedded in $R^n$. The dual torus on 
which the Nahm transformed gauge field lives is $\hat T^n=R^n/\hat\Lambda$. 
Hence, for $n\neq 4$ there is a reduction in dimension. The extreme case is 
when $n=0$, which reduces to the algebraic ADHM~\cite{ADHM} construction on 
$R^4$. The case $n=1$ has also led to considerable progress in demonstrating 
that instantons at finite temperature (calorons) have monopoles as 
constituents~\cite{KrvB,LeLu}. Important pratical use of the Nahm 
transformation stems from the fact that the charge one solutions are mapped 
to self-dual Abelian fields. 

Recently Gonz\'alez-Arroyo~\cite{TNT} has constructed the Nahm 
transformation in the presence of twisted boundary conditions for $T^4$. 
One interprets twisted boundary conditions as ``half-period'' conditions (for 
$SU(2)$ a quite appropriate terminology), applying the Nahm transformation to 
the gauge fields on the (smallest) extended torus with periodic boundary 
conditions. One subsequently looks for ``half-periods'' on the dual torus. 
This elegant construction will lead to important new insights, but increases 
the topological charge of the periodic solutions. It forces one to deal with 
Nahm gauge fields that are non-Abelian, not giving an obvious simplification 
that is likely to lead to a handle on an analytic construction.

Here we will be concerned with $T^3\times R$, for which the Nahm transformation
was introduced a few years ago~\cite{Buck}, and which is relevant for the 
Hamiltonian formulation of gauge theories in a finite volume with periodic 
boundary conditions (for a recent review addressing the dynamical issues see 
ref.\cite{Edin}). In this Letter we will present the analytic solution for 
the gauge field obtained after applying the Nahm transformation to a charge 
one instanton, in terms of its flat connections at $t\rightarrow\pm\infty$. 
That the instanton has to approach a flat connection in these limits is a 
simple consequence of the requirement of finite action. 

\section*{}
2. We first present the formalism, being rather brief and referring the
interested reader to refs.~\cite{BrvB,Nahm,Buck} for details. A self-dual 
$U(N)$ gauge field $A(x)=A_\mu(x)dx_\mu$ (with $A^\dagger=-A$) is defined 
on $R^4/\Lambda$ by 
\be
A(x+\lambda)=g_\lambda(x)(A(x)+d)g_\lambda^\dagger(x),\quad
\lambda\in\Lambda.
\ee
It is made into a family of self-dual gauge fields by adding a (flat) constant 
Abelian connection, $A_z(x)=A(x)+2\pi i d(z\cdot x)$. Note that the differential
is with respect to $x$ only, $d(z\cdot x)=z\cdot dx=z_\mu dx_\mu$, and an 
identity matrix, $I_N$, in the algebra of $U(N)$ is implicit. Even though the 
curvature (field strength) is independent of $z$, its dependence cannot be 
gauged away since the appropriate Abelian gauge transformation $g(x)=\exp(2\pi i
z\cdot x)$ is not periodic, except when $z\in\hat\Lambda$. This shows that $z$
can be considered to live on the dual space $R^n/\hat\Lambda$. The reduction in 
dimension alluded to above occurs since for non-compact directions the relevant
components of $z$ {\em can} be gauged away. Equivalently, non-compact directions
can be interpreted as having infinite periods, which under the duality are 
mapped to zero periods, removing the dependence on the dual coordinate.

The Nahm transformation involves the zero-modes of the Weyl equation, of which 
there are as many as the charge ($k$) of the gauge field 
\bea
D_z\Psi_z(x)&=&\sigma_\mu D_\mu(A_z)\Psi_z(x)=\sigma_\mu(\partial_\mu+
A_\mu(x)+2\pi i z_\mu)\Psi_z(x)=0,\nonumber\\
\Psi_z(x+\lambda)&=&g_\lambda(x)\Psi_z(x),\quad\lambda\in\Lambda,
\eea
where $\sigma_\mu$ form a basis of unit quaternions ($\sigma_0=I_2$ and 
$-i\sigma_j=\tau_j$ the Pauli matrices). As $\Psi_z$ is in the fundamental 
representation of the gauge group we can not allow for twisted boundary 
conditions, which require the center of the gauge group to act trivially. As 
mentioned above one can enlarge the periods~\cite{TNT} to deal with twisted 
boundary conditions. Here we will instead consider only twisted boundary 
conditions in the non-compact directions, where the action of the center of 
the gauge group is trivialised~\cite{Buck} due to Weyl fermions vanishing 
asymptotically, so as to ensure normalisability.

The Nahm connection is given in terms of the normalised zero-modes by
\be
\hat A^{ij}(z)=\int d^4x~\Psi_z^{(i)}(x)^\dagger\frac{\partial}{\partial z_\mu}
\Psi_z^{(j)}(x) dz_\mu.
\ee
It is not difficult to show that this is a $U(k)$ connection on $R^4/\hat
\Lambda$ and using the family index theorem one concludes~\cite{BrvB} the
topological charge of the Nahm gauge field to be $N$. The index theorem 
relates the difference of the number of zero-modes with opposite chirality to 
the topological charge, that is ker($D_z$)--coker($D_z$)=ker($D_z$)--ker($
D_z^\dagger$) has dimension $k$. Additional (i.e. non-generic) zero-modes 
are therefore detected as zero-modes of 
\be
D_zD_z^\dagger=-D_\mu^2(A_z)-\half\bar\eta_{\mu\nu}F_{\mu\nu}(x),
\ee
where 
$\bar\eta_{\mu\nu}=\sigma^{\hphantom{\dagger}}_{[\mu}\sigma^\dagger_{\nu]}$
is the anti-self dual 't Hooft tensor. For self-dual fields we 
have that $D_zD_z^\dagger=-D^2_\mu(A_z)$, which commutes with $\sigma_\mu$, 
from which Nahm derived the remarkable result that the Nahm gauge field is 
self-dual as well. For $T^4$ this is most easily demonstrated, as the manifold 
has no boundaries. Technically one requires the absence of flat 
factors~\cite{DoKr} to ensure that ker($D_z^\dagger$) is trivial.

For a non-compact manifold, applying index theorems requires some care, but in 
principle $\hat A(z)$ is well defined as long as dim ker($D_z^\dagger$)=0, and
one finds~\cite{Buck}
\bea
\hat F_{\mu\nu}^{ij}(z)&=&8\pi^2\int d^4xd^4x'~\Psi^{(i)}_z(x)^\dagger
\eta_{\mu\nu}G_z(x,x')\Psi_z^{(j)}(x')\\
&+&4\pi i\oint\frac{\partial\Psi^{(i)}_z(x)^\dagger}{\partial z_{[\mu}}
\eta_{\nu]\alpha}\left(G_z\Psi_z^{(j)}\right)
(x)\,d^3_\alpha x,\nonumber
\eea
where $\eta_{\mu\nu}=\sigma^\dagger_{[\mu}\sigma^{\hphantom{\dagger}}_{\nu]}$
is the self-dual 't Hooft tensor and $G_z$ is the Greens function 
for $-D_\mu^2(A_z)$.

\section*{}
3. On $T^4$, applying the Nahm transformation again brings one back to the 
original solution. In other cases one needs to modify the second, or inverse,
Nahm transformation. The boundary terms are particularly important in the case 
of instantons on $R^4$, leading to the ADHM construction~\cite{ADHM} for 
reconstructing the original gauge field. The modification corrects for the 
fact that $\hat A(z)$ is no longer self-dual due to the boundary terms. However,
boundary terms only occur at a finite number of isolated points, and can be 
expressed in terms of delta functions (excluding the situation of $R^4$ where 
the dual space is reduced to a single point). The singularities are fixed by 
the asymptotic holonomies (the Polyakov loops). For $T^3\times R$ there are 
two disconnected asymptotic regions and we specify 
\be
P_\pm(\vec n,z)\equiv\lim_{t\rightarrow\pm\infty}P\exp(\int_{C(\vec n)}A_z(x)),
\ee
where $\vec n$ is the number of windings for each direction on $T^3$, specifying
the homotopy type of the curve $C(\vec n)$. The $N$ eigenvalues of $P_\pm
(\vec n,z)$ are given by $\exp(2\pi i(\vec\omega_\pm^j+\vec z)\cdot\vec n)$.
For $SU(N)$ gauge fields one has in addition $\sum_j\vec\omega^j_\pm=\vec 0$. 

It is now easily seen~\cite{Buck}, when all eigenvalues are {\em unequal} 
to one, that a Weyl zero-mode decays as $\exp(\mp t M_\pm(z))$, where 
$M_\pm(z)$ is the mass-gap of the Weyl equation reduced to $T^3$ for 
$t\rightarrow\pm\infty$, with
\be
M_\pm(z)={\rm min}\{2\pi|\vec\omega_\pm^j+\vec z+\vec p|;j=1,\cdots,N;
\vec p\in Z^3\}.
\ee
Indeed, $M_\pm(z)$ vanishes whenever $P_\pm(\vec n,z)$ has a unit eigenvalue
for all $\vec n$. Only for those cases the boundary terms arising from a 
partial integration in computing $\hat F$ can be non-vanishing.

Outside of the singularities the field is self-dual and $\hat E_i(z)=\hat
B_i(z)$. As $\hat A(z)$ is independent 
of $z_0$ we have $\hat E_i(z)=i\partial_i\hat A_0(z)$. We extract a factor $i$,
to define $\hat E_i(z)$ and $\hat B_i(z)(=-\frac{i}{2}\varepsilon_{ijk}\hat 
F_{jk}(z))$ as real fields. Assuming other than periodic boundary conditions, 
near one of the singularities integration by steepest descent yields 
$\hat A_0(z)\rightarrow\pm i q/M_\pm(z)$, where $q$ is a positive constant. 
This means that the singularities act as point sources with charges $\pm q$. 
Since self-duality implies that the (Maxwell) field equations are satisfied, 
the exact solution is found by performing the sum over periods for these point 
charges. But we can say more. For the gauge field to be well-defined outside 
of the singularities, charge quantisation should be enforced and one concludes 
that $q/\pi$ has to be an integer, but generically $q=\pi$. This ensures the 
magnetic sources are those of Dirac monopoles with unobservable Dirac strings. 
We will now demonstrate this on the basis of a Berry phase type argument.

\section*{}
4. We restrict ourselves to the case of twisted boundary conditions in the 
time direction. For ease of notation we take all three periods equal to one 
(generalisation to another torus is straightforward) and we consider 
$T^3\times R$ as the limit for $T\rightarrow\infty$ of $T^3\times[0,T]$. The 
twist can be implemented by choosing~\cite{THotw,CMP} the gauge field to be 
periodic in the spatial directions, whereas the gauge field at $t=T$ is 
related to the one at $t=0$ by a gauge transformation $g(\vec x)=g_k(\vec x)
g_{\vec k}(\vec x)$. Here $g_k(\vec x)$ is a periodic gauge transformation 
with winding number $k$ and $g_{\vec k}(\vec x)=\exp(2\pi i\vec x\cdot\vec k
\Theta)$, with $\Theta$ ($=\half\tau_3$ for $SU(2)$) such that $g_{\vec k}(\vec
n)=\exp(2\pi i\vec k\cdot\vec n/N)\in Z_N$ for $\vec n\in Z^3(=\!\Lambda)$. For
finite $T$ and $\vec k \neq\vec 0$ mod $Z_N^3$ existence of a $4Nk$ parameter 
set of solutions is guaranteed~\cite{Bra}. Taking $T\rightarrow\infty$ yields 
solutions on $T^3 \times R$. With twisted boundary conditions $P_+(\vec n,z)=
\exp(2\pi i\vec k\cdot\vec n)P_-(\vec n,z)$, relating the singularities 
discussed above. 

Consider $T$ finite and add an Abelian background gauge field, whose flux 
compensates for the twist~\cite{LuVe}. The price one pays is that the 
$U(N)$ gauge field will in general no longer be self-dual. We introduce the 
periods $L_\mu$ and the antisymmetric twist tensor $n_{\mu\nu}$, where in the 
case at hand we would have $L_i=1$, $L_0=T$, whereas $n_{0i}=k_i$. One defines
\be
\bar A_\mu=\pi i n_{\mu\nu}x_\nu I_N/(N L_\mu L_\nu),\quad
\bar F_{\mu\nu}=-2\pi i n_{\mu\nu} I_N/(N L_\mu L_\nu).
\ee
In terms of the curvature two-form $\bar\Omega=\half\bar F_{\mu\nu} 
dx_\mu\wedge dx_\nu$, the first Chern class is given by
\be
c_1=Tr\bar\Omega/(2\pi i).
\ee
The Pontryagin index for the $U(N)$ bundle $A+\bar A$ is now
\be
P_t=(8\pi^2)^{-1}\int Tr(\Omega+\bar\Omega)\wedge(\Omega+\bar\Omega)=
(8\pi^2)^{-1}\int Tr\Omega\wedge\Omega+Tr\bar\Omega\wedge\bar\Omega,
\ee
where $\Omega$ is the curvature two-form of the original (self-dual) $SU(N)$ 
connection, satisfying $Tr\Omega=0$. From ref.~\cite{CMP} we find 
$P=\nu-(N-1)\pf(n)/N$, with $\nu$ integer, such that $P_t=P+\bar P=
P-\int c_1\wedge c_1/(2N)=P-\pf(n)/N$. Thus, $P_t=\nu-\pf(n)$ is always integer
as required for $U(N)$ vector bundles. For the case at hand, with only twist 
in the time direction, $\pf(n)=0$ and the topological charge is not affected 
by adding the twist compensating Abelian background field.

The Nahm transformation maps this to a bundle with rank $P_t$, charge $N$ and 
first Chern class $\int_{T^4}(dz_\mu\wedge dx_\mu)^2\wedge c_1$ (for the 
precise formulation see ref.~\cite{BrvB}). Consider now the case of 
topological charge one. Assuming the Weyl cokernel to be trivial for all $z$, 
we get a (non-selfdual) $U(1)$ connection with charge $\hat P=N$. But for an 
Abelian connection we also have $\hat P=-\half\int\hat c_1\wedge\hat c_1=-\pf(n)
=0$. So there must be values of $z$ for which the Weyl cokernel is non-trivial.
We have
\be
D_zD_z^\dagger=-D_\mu^2(A_z+\bar A)+i\pi n_{\mu\nu}\bar\eta_{\mu\nu}
I_N/(NL_\mu L_\nu)=-D_\mu^2(A_z+\bar A)+\vec H\cdot\vec\tau,
\ee
with $H_k=2\pi[n_{0k}/(NL_0L_k)-\half\epsilon_{ijk}n_{ij}/(NL_iL_j)]$, or for  
the case at hand $\vec H=2\pi\vec k/(NT)$. We note that $D_zD_z^\dagger$ is 
positive, but it may vanish. Using $\vec H\cdot\vec\tau$ has eigenvalues 
$\pm|\vec H|$, one easily finds this to be the case if and only if the 
positive function $f(z)\equiv\lambda_0(z)-|\vec H|$ vanishes, where 
$\lambda_0(z)$ is the lowest eigenvalue of $-D_\mu^2(A_z+\bar A)$. With
$\Phi_z(x)$ its normalised eigenvector, we have $f(z)=\int d^4x~\Phi_z^\dagger
(x)D_zD_z^\dagger\Phi_z(x)$ such that $\partial^2f(z)/\partial z_i\partial z_j=
8\pi^2\delta_{ij}$ at the points where $f(z)$ vanishes. Considering the fact 
that $f(z)$ is a smooth and positive function of $z$, the zeros are generic 
and cannot bifurcate in zeros of lower order. 

As long as $T$ is finite we can use the index theorem and conclude that at 
exactly the same points where dim ker($D_z^\dagger$) jumps from zero to one, 
dim ker($D_z$) has to jump from one to two. Since $D_z$ depends smoothly on 
$z$, this necessarily describes the case of a generic level crossing. The only 
unusual feature is that the ``Hamiltonian'' is arranged so as to vanish 
exactly for one of the "adiabatic" eigenstates. The resulting Berry 
potential~\cite{Berry} associated to the isolated crossing corresponds 
precisely to the spatial components of the Nahm connection, eq.(3). Due to 
the topological nature of the Berry phase we can immediately conclude that 
the level crossing acts as the source of a Dirac monopole with the appropriate 
charge quantisation, enforcing $q=\pi$. It is clear that this assignment is 
independent of $T$ and this fixes the charges for the case of $T^3\times R$. 

Note that $\bar A\rightarrow 0$ for $T\rightarrow\infty$. We reiterate that 
under this limit the total action stays fixed to $8\pi^2$, as dictated by the 
unit topological charge, and the fields are forced to decay to flat connections
at both ends, thereby dictating the location of singularities that act as point
sources. We note that for twisted boundary conditions one can put 
$\vec\omega_+^j=\vec\omega_-^j+\vec k/N$~mod$Z^3$. Generically there are $N$ 
sources with charge $q=\pi$ and $N$ sources with the opposite charge. Higher 
charges appear only in case some of the $\vec\omega_\pm^j$ coincide. The 
enlarged subgroup that leaves the holonomies invariant leads to appropriate 
additional zero-modes for $D_z^\dagger$. 

\section*{}
5. The Nahm connection on $\hat T^3$ is uniquely determined by the point 
charges we described above, 
\be
\hat A_0=\frac{i}{2}\sum_{\vec n\in Z^3}\sum_{j=1}^N \left(|\vec\omega_+^j+
\vec z+\vec n|^{-1}-|\vec\omega_-^j+\vec z+\vec n|^{-1}\right).
\ee
One difficulty is to evaluate the sum over the periods, as it formally 
diverges. This can be achieved in terms of lattice sum techniques based on 
resummations~\cite{NdW}. Quite fortunately this problem was already tackled 
long ago in evaluating the one-loop effective potential for constant Abelian 
gauge fields on $T^3$, $V_1(\vec C)=2\sum_{\vec n\in Z^3}|2\pi\vec n+\vec C|$. 
In terms of $W(\vec C/2\pi)\equiv\half\pi\Delta V_1(\vec C)$ one easily finds
\be
\hat A_0(\vec z)=\frac{i}{2}\sum_{j=1}^N\left(W(\vec z+\vec\omega_+^j)-
W(\vec z+\vec\omega_-^j)\right),
\ee
where the rapidly converging expression for $W(\vec z)$ can be taken 
from eq. (A.10) of ref.~\cite{APh},
\be
W(\vec z)=-1+\sum_{\vec n\neq\vec 0}\frac{e^{-\pi\vec n^2}}{\pi\vec n^2}
\cos(2\pi\vec n\cdot\vec z)+\sum_{\vec n}\frac{{\rm erfc}(|\vec n+\vec z|
\sqrt{\pi})}{|\vec n+\vec z|}
\ee
It can be shown~\cite{APh} that $\Delta W(\vec z)=-4\pi(\delta(\vec z)-1)$, 
where $\delta(\vec z)$ is the periodic delta function, such that indeed
\be
\partial_i\hat E_i(\vec z)=2\pi\sum_{j=1}^N\left(
\delta(\vec z+\vec\omega_+^j)-\delta(\vec z+\vec\omega_-^j)\right).
\ee

It is still a formidable task to reconstruct from this explicit expression for 
the Nahm connection the original non-Abelian gauge field on $T^3\times R$. 
This requires, like for the simpler case of the calorons~\cite{KrvB}, the 
formulation of a modified Nahm transformation, dealing with the singularities 
to which violations of self-duality are restricted. Nevertheless, given the 
existence of solutions with twisted boundary conditions interesting 
conclusions can be drawn. Up to an overall constant, related to the position 
of the instanton on $T^3\times R$, $\hat A(\vec z)$ is determined uniquely by 
the eigenvalues of the holonomies. These holonomies, when taking the limit 
$T\rightarrow\infty$ (unlike in the case of the calorons) arise from the 
properties of the solutions on $T^4=T^3\times[0,T]$, and are thus part of the 
gauge invariant moduli. Together with the position of the charge one instanton, 
these account for the $3(N-1)+4$ parameters of the gauge invariant moduli space.
The holonomy breaks the gauge group to $U(1)^{N-1}$, accounting for $N-1$ 
additional parameters that are part of the moduli space of framed instantons, 
which has dimension $4N$, as is appropriate for the torus and this leaves no 
room for a scale parameter of the instanton. We thus conclude that the size 
of the instanton is related to the holonomies, something that was 
conjectured~\cite{Buck} on the basis of numerical studies~\cite{GG} and 
in direct analogy with the situation for instantons on the cylinder for 
the $O(3)$ model~\cite{Jer}. 

For $SU(2)$ it was indeed observed that the largest instanton, the one that 
described tunnelling through the lowest barrier (sphaleron), is associated 
to $\vec k=(1,1,1)$ and all holonomies equal to $\pm 1$. This corresponds 
to $\vec\omega_-$ and $\vec\omega_+(=\half\vec k+\vec\omega_-)$ separated on 
$\hat T^3$ over the maximal distance possible. On the other hand, when the 
trace of the holonomy vanishes, $\pm\vec\omega=\quart\vec k$, the twisted 
boundary conditions are compatible with periodic boundary conditions, and 
indeed $\vec\omega_-$ and $\vec\omega_+$ become equal and $\hat A(\vec z)$ 
will vanish (apart from a trivial constant). In that case, like in the analysis
for the caloron~\cite{KrvB}, associating the Nahm and ADHM formulation by 
Fourier transformation the solution becomes expressible in terms of the 
't Hooft ansatz~\cite{THoAn}, $A_\mu=\half\bar\eta_{\mu\nu}\partial_\nu\log
\phi(x)$, with $\phi(x)=\rho^{-2}+\sum_{\vec n}[t^2+(\vec x+\vec n)^2]^{-1}$. 
Resummation of the lattice sum is easily performed, but positivity of $\phi(x)$
is seen to force $\rho$ to zero. 

\section*{}
6. It is tempting to conjecture on the basis of our results for the Nahm 
connection that solutions will exist for open boundary conditions,
in which case the holonomies at both ends are not related\footnote{The number 
of gauge invariant parameters describing such solutions would be $6(N-1)+4$.
As these solutions cannot be compactified to $T^4$, there need be no 
conflict with the standard result on a compact manifold~\relax{\cite{DoKr}}.}.
If true, we can obtain periodic boundary conditions as a limit from the 
one with open boundary conditions, implying $\hat A(\vec z)$ to approach 
a constant, and one would as above conclude that this forces the size of 
the instanton to zero.

It is amusing in the light of this to note that, when extending the
caloron construction~\cite{KrvB} based on Fourier transformation of 
the ADHM formulation in an obvious way to $T^3\times R$ (e.g. see
eq.~(40) in ref.~[8b]), one finds
\be
\tau_j(\hat E_j(\vec z)-\hat B_j(\vec z))=\vec a\cdot\vec\tau\left(
\delta(\vec z-\vec\omega)-\delta(\vec z+\vec\omega)\right). 
\ee
where for simplicity we only considered $SU(2)$. Here the direction of 
$\vec a$ is related to the common gauge orientation of the holonomies and its 
length is related to the square of the scale parameter that appears in the 
ADHM construction. Due to the vectorial nature of the singularity it is 
natural to assume the gauge field is described by an electric-magnetic 
dipole.  Remarkably, it is well known in the theory of classical 
Electrodynamics (e.g. see ref.~\cite{Jac}) that for dipoles 
\be
\vec E(\vec x)=\frac{3\vec x(\vec p\cdot\vec x)-\vec p(\vec x\cdot\vec x)}{
|\vec x|^5}-\frac{4\pi}{3}\vec p\delta(\vec x),\quad
\vec B(\vec x)=\frac{3\vec x(\vec m\cdot\vec x)-\vec m(\vec x\cdot\vec x)}{
|\vec x|^5}+\frac{8\pi}{3}\vec m\delta(\vec x).
\ee
These delta functions differ between electric and magnetic dipoles, as the first
comes from two approaching point charges and the second from a shrinking current
loop. Thus, with $\vec p=\vec m$, one finds $\vec B(\vec x)-\vec E(\vec x)=4\pi
\vec m\delta(\vec x)$, precisely of the required form. The appropriate solution
for $SU(2)$ is described by two ``dyonic'' dipoles of opposite strength located
at $\vec z=\pm\vec\omega$. Outside the singularities this can easily be 
expressed in terms of the lattice sums we defined before,
\be
\hat B_i(\vec z)=
\hat E_i(\vec z)=\frac{a_j}{4\pi}\frac{\partial^2}{\partial z_i\partial z_j}
\left(W(\vec z+\vec w)-W(\vec z-\vec w)\right).
\ee
Indeed, resolving the quadratic ADHM constraint~\cite{ADHM} and explicit 
Fourier transformation reproduces this result, including the appropriate 
delta functions that violate the self-duality~\cite{Kra}. 

Nevertheless, as we cannot argue for the existence of solutions with periodic 
boundary conditions on $T^3\times R$, it may be that this dipole solution to the
Nahm equations is not realised, except for $\vec a\rightarrow\vec 0$, implying 
the size of the instanton to go to zero. The dipole approximation obtained
from open boundary conditions, letting $\vec\omega_+$ tend to $\vec\omega_-$ 
to approach periodic boundary conditions, indeed leads to vanishing dipole
moments (since the charge is fixed). This conclusion can only be avoided in 
case {\em no} solutions with open boundary conditions exist.

\section*{}
7. In conclusion, we have shown that one can extract analytic results for the 
Nahm transformation of the charge one instanton on $T^3\times R$, which 
provides interesting information on the parameters that describe
the solutions. The situation is quite similar to that for the $O(3)$
model on the cylinder. It will be interesting to be able to demonstrate 
the existence of solutions with open boundary conditions. 
However, the biggest challenge remains the formulation of the inverse Nahm 
transformation, which requires us to study the (modified) Weyl equation in 
a lattice of ``dyonic'' charges. 

Numerical studies of the Nahm transformation~\cite{GAP} have been implemented, 
and may well play a role in analytically addressing these issues. Preliminary 
results~\cite{Pena} relevant for the case studied here are very encouraging 
and stimulating. Also the deformation of the Nahm transformation to the 
noncommutative torus~\cite{NCT} and its M-theory compactifications~\cite{HoVe} 
could perhaps provide insight in the problem addressed here.

\section*{Acknowledgements}

Stimulating discussions with Margarita Garc\'{\i}a P\'erez, Tony 
Gonz\'alez-Arroyo, Thomas Kraan and Carlos Pena are gratefully acknowledged.
Most of the ideas presented here were initiated while I was visiting the 
Newton Institute during the first half of 1997, giving me another opportunity 
to thank the staff for their hospitality and the participants to the programme 
on ``Non-perturbative Aspects of Quantum Field Theory'' for contributing to a 
stimulating environment.

\end{document}